\documentclass[twocolumn,aps,prb,showpacs,amsmath,amssymb,superscriptaddress,bibnotes,longbibliography]{revtex4-2}

\usepackage{hyperref}
\usepackage[usenames,dvipsnames]{color}
\usepackage{amsmath}
\usepackage{amssymb}
\usepackage{amsthm}
\usepackage{array}
\usepackage{graphicx}
\usepackage{epstopdf}
\usepackage{todonotes}
\usepackage[normalem]{ulem}
\usepackage{verbatim}
\usepackage{stmaryrd}
\usepackage{lipsum}
\usepackage{mathtools}
\usepackage{xcolor}
\usepackage{diagbox}
\usepackage{tabularx}
\normalsize
\usepackage{booktabs}

\usepackage[export]{adjustbox}

\newcommand{\T}{\mathcal{T}}

\newcommand{\bG}{\mathbf{G}}
\newcommand{\bSigma}{\mathbf{\Sigma}}
\newcommand{\bS}{\mathbf{S}}
\newcommand{\bh}{\mathbf{h}}
\newcommand{\bF}{\mathbf{F}}
\newcommand{\bP}{\mathbf{P}}
\newcommand{\bM}{\mathbf{M}}
\newcommand{\bX}{\mathbf{X}}
\newcommand{\bY}{\mathbf{Y}}
\newcommand{\bR}{\mathbf{R}}
\newcommand{\bbf}{\mathbf{f}}
\newcommand{\bx}{\mathbf{x}}
\newcommand{\by}{\mathbf{y}}
\newcommand{\bz}{\mathbf{z}}
\newcommand{\be}{\mathbf{e}}
\newcommand{\bk}{\mathbf{k}}
\newcommand{\bbr}{\mathbf{r}}

\begin{document}
\title{Equivariant neural network for Green's functions of molecules and materials}

\author{Xinyang Dong}
\email{xinyang.dongxy@gmail.com}
\affiliation{AI for Science Institute, Beijing 100080, China}
\author{Emanuel Gull}
\affiliation{Department of Physics, University of Michigan, Ann Arbor, MI 48109, USA}
\author{Lei Wang}
\affiliation{Beijing National Laboratory for Condensed Matter Physics and Institute of Physics, \\Chinese Academy of Sciences, Beijing 100190, China}
\affiliation{Songshan Lake Materials Laboratory, Dongguan, Guangdong 523808, China}

\date{\today} 

\begin{abstract}
     The many-body Green's function provides access to electronic properties beyond density functional theory level in ab inito calculations. 
     In this manuscript, we propose a deep learning framework for predicting the finite-temperature Green's function in atomic orbital space, aiming to achieve a balance between accuracy and efficiency.
     By predicting the self-energy matrices in Lehmann representation using an equivariant message passing neural network, our method respects its analytical property and the $E(3)$ equivariance. 
     The Green's function is obtained from the predicted self-energy through Dyson equation with target total number of electrons.
     We present proof-of-concept benchmark results for both molecules and simple periodic systems, showing that our method is able to provide accurate estimate of physical observables such as energy and density of states based on the predicted Green's function.
\end{abstract}

\maketitle
\makeatletter
\let\toc@pre\relax
\let\toc@post\relax
\makeatother

\section{Introduction}

The single-particle Green's function plays a fundamental role in the computational study of quantum field theories in condensed matter physics, quantum chemistry, and material science.
It provides in particular information about the single-particle excitation spectrum, which can be  compared to scanning tunneling microscopy and angle-resolved photoemission spectroscopy experiments.
In recent years, rapid development in ab initio theory \cite{Dahlen_2005,Zgid_2011,Lan_2015,Rusakov_2019,Zhu_2020} and the numerical implementation \cite{Shinaoka_2017,Gull_2018,Iskakov_2019,Shee_2019,Dong_2020,Backhouse_2020,Li_2020,Bintrim_2021,Dong_2022,Yeh_2022} of Green's function methods has enabled systematic calculations of interacting quantum many-body systems.

The field theory formulation provides a theoretically rigorous view of the finite-temperature physics of interacting quantum systems, complementary to results from ground state methodologies such as the density functional theory (DFT). However,
computing Green's functions is in general orders of magnitude more expensive than a DFT calculation, limiting the methodology to small systems.

This motivates research into the application of data-driven machine learning approaches to quantum field theories. Such methods balance accuracy with efficiency, and prior work has shown considerable success.
For instance, Refs.~\onlinecite{Arsenault_2014,Sheridan_2021,Sturm_2021} developed  machine learning models to predict the Green's function of the single-site Anderson impurity model, serving as impurity solvers for the dynamical mean field theory (DMFT). 
Ref.~\onlinecite{Venturella_2024} employed  Kernel Ridge Regression to predict self-energies and spectral functions of realistic systems starting from a mean-field Hartree-Fock solution.
Still, the power of state-of-the-art deep learning models of finite-temperature field theories when applied to realistic systems has yet to be demonstrated.

In this manuscript, we propose a deep learning framework for predicting the many-body Green's function and self-energy for both molecules and periodic systems in atomic orbital space.
We employ a message passing neural network \cite{Gilmer_2017,Battaglia_2018} that maps atomic configurations to matrices, in analogy to a framework that successfully predicts  DFT Hamiltonians \cite{Li_2022,Su_2023,Unke_2021,Gong_2023,Yang_2023,Yu_2023}.
To achieve better accuracy and data efficiency in the training process, we use an equivariant setup as in Refs.~\onlinecite{Thomas_2018,Unke_2021,Brandstetter_2022,Geiger_2022,Batzner_2022,Musaelian_2023,Gong_2023,Yang_2023,Yu_2023}. 
By constructing the fundamental self-energy matrices in the Lehmann representation using the equivariant features, we ensure the fulfillment of their analytical properties by construction. 

This paper is organized as follows: 
In section \ref{sec:method}, we introduce the theory of finite temperature Green's function (sec.~\ref{sec:finite_gf}), the equivariant message passing neural network (sec.~\ref{sec:equiv_mpnn}), and how we employ the neural network to predict self-energies and Green's functions (sec.~\ref{sec:gf_nn},\ref{sec:workflow}).
In section \ref{sec:results}, we present results of proof-of-concept benchmarks for both molecules (sec.~\ref{sec:mol}) and periodic systems (sec.~\ref{sec:pbc}).
Section \ref{sec:conclusion} provides conclusion and outlook.

\section{Method} \label{sec:method}

\subsection{Green's function formalism} \label{sec:finite_gf}

Within the Born-Oppenheimer approximation and in the absence of relativistic effects, the second-quantized Hamiltonian of realistic systems can be written as \cite{Negele_1988,Szabo_1996}
\begin{align}
    H = \sum_{ij} \sum_{\sigma} h_{ij} c_{i\sigma}^{\dagger} c_{j \sigma} + \frac{1}{2} \sum_{ijkl} \sum_{\sigma \sigma'} U_{ijkl} c_{i\sigma}^{\dagger} c_{k\sigma'}^{\dagger} c_{l\sigma'} c_{j\sigma} \, , \label{eq:Hamiltonian_operator}
\end{align}
where $c_{i\sigma}^{\dagger}$($c_{i\sigma}$) are the creation (annihialation) operators with orbital index $i$ and spin index $\sigma$, $h_{ij}$ is the one-electron kinetic and electron-nuclei integral, and $U_{ijkl}$ the Coulomb repulsion integral. 
The atomic orbitals $g_i(\bbr)$ considered in this work may be non-orthogonal, defining an overlap matrix in  orbital space \cite{Szabo_1996},
\begin{align}
    S_{ij} = \int d\bbr g_i^*(\bbr) g_j(\bbr) \, . 
\end{align}
We will use $i$ as spin-orbital index to omit the explicit spin index $\sigma$ and use bold symbols for matrices in spin-orbital space in the rest of this paper.

To compute physical properties of an electron system, we introduce the  single-particle finite temperature Green's function \cite{Negele_1988,Mahan_2012}
\begin{align}
    G_{ij}(\tau) = -\langle \T c_{i}(\tau) c_{j}^\dagger(0) \rangle \, ,
\end{align}
where $\tau \in [0, \beta]$ is the imaginary time, $\beta = \frac{1}{k_B T}$ the inverse temperature, and $\mathcal{T}$ the time ordering operator \cite{Negele_1988} .
The imaginary time Green's function $\bG(\tau)$ corresponds to a frequency space or Matsubara Green's function with the transform 
\begin{align}
    \bG(i\omega_n) = \int_0^{\beta} d\tau \, \bG(\tau) e^{i\omega_n \tau} \, ,
\end{align}
where the fermionic Matsubara frequencies $\omega_n$ are defined as $\omega_n = (2n + 1)\pi/\beta$, $n \in \mathbb{Z}$.
The connection between the non-interacting Green's function $\bG_0$  and the full Green's function $\bG$ is given by the Dyson equation
\begin{align}
    \bG(i\omega_n) = \bG_0(i\omega_n) + \bG_0(i\omega_n) \bSigma(i\omega_n) \bG(i\omega_n) \, , \label{eq:dyson}
\end{align}
where $\bG_0(i\omega_n) = [(i\omega_n + \mu) \bS - \bh]^{-1}$, with $\mu$ the chemical potential, $\bh$ as in Eq.~\ref{eq:Hamiltonian_operator} and $\bSigma(i\omega_n)$ the Matsubara frequency self-energy which is a function of the full Green's function $\bSigma \equiv \bSigma[\bG]$.
The self-energy can be split into two parts,
\begin{align}
    \bSigma[\bG](i\omega_n) = \bSigma^\text{(HF)}[\bG] + \tilde{\bSigma}[\bG] (i\omega_n) \, ,
\end{align}
where $\bSigma^\text{(HF)}[\bG]$ is the static Hartree-Fock (HF) self-energy and $\tilde{\bSigma}[\bG] (i\omega_n)$ denotes the frequency-dependent dynamical self-energy. Usually $\bSigma^\text{(HF)}[\bG]$ is combined with the one-body integral $\mathbf{h}$ into the so-called Fock matrix $\bF = \bh + \bSigma^\text{(HF)}$.
The total energy of the system is given by
\begin{align}
    E_\text{tot} 
    &= \frac{1}{2} \text{Tr}[\mathbf{P}(\bh + \bF)] + \frac{1}{2}\text{Tr}[\tilde{\bSigma} \ast \bG] + E_{nn} \, ,
    \label{eq:e_tot}
\end{align}
where $E_{nn}$ is the nuclei-nuclei Coulomb energy, $\bP = \bG(\tau=0^{-})$ is the density matrix, $\ast$ is the imaginary time convolution operator, and the trace is defined as $\text{Tr}[\mathbf{A}] = -\sum_i A_{ii} (\beta^{-})$.

The Green's function can be analytically continued from the Matsubara frequencies to the entirety of the complex plane, and will be analytic in the upper half of the complex plane. The limit of the Green's function taken towards the real frequency axis corresponds to the so-called retarded Green's function. It can be obtained from Matsubara data with numerical analytical continuation and directly yields the spectral function (DOS) of the system,
\begin{align}
    A(\omega) = -\frac{1}{\pi} \text{Im}(\text{Tr}[\bS \bG(\omega)]) \, .
    \label{eq:DOS}
\end{align}
Refs.~\onlinecite{Stefanucci_2013,Luttinger_1960,Dahlen_2006,Phillips_2014, Rusakov_2016,Welden_2016,Iskakov_2019,Yeh_2022} contain further references and detailed explanations of the finite-temperature formalism and its numerical implementation.

\subsection{Equivariant message passing neural networks} \label{sec:equiv_mpnn}

In graph neural networks (GNN) or message passing neural networks (MPNN), atomic structures of isolated molecules or periodic solids are represented by nodes and edges, where nodes indicate atoms and edges demonstrate the connection between atom pairs \cite{Gilmer_2017,Battaglia_2018}. 
In the `message passing' process, starting from the initial element based embedding such as the one-hot encoding of nuclear charge, the feature vector $\bbf_i$ associated with node $i$ is iteratively updated through convolutions with its neighbors based on their features $\bbf_j$ and distances $\mathbf{r}_{ij}$.

The node features of the graph can be used to construct desired physical quantities such as the inter-atomic potential \cite{Batzner_2022,Musaelian_2023} or DFT (tight-binding) hamiltonian \cite{Unke_2021,Li_2022,Gong_2023,Su_2023,Yang_2023,Yu_2023}.
The GNN (MPNN) setup can be either rotational invariant or equivariant. 
Since the objects we are interested in are matrices in atomic orbital space which are equivariant under rotations, we choose to use the equivariant message passing neural network.

The core operation in equivariant neural network architecture is the tensor product operation that couples two representations in an equivariant way \cite{Geiger_2022}
\begin{align}
    &\bz^{(l_3)} = \bx^{(l_1)} \otimes \by^{(l_2)} \, ,
    \\
    &z_{m_3}^{l_3} =
    \sum_{m_1=-l_1}^{l_1} \sum_{m_2=-l_2}^{l_2} C_{m_3, m_2, m_1}^{l_3, l_2, l_1} x_{m_1}^{l_1} y_{m_2}^{l_2} \, ,
\end{align}
where $C$ denotes the Clebsch-Gordan (CG) coefficients, $l \in \mathbb{N}$ are angular momentum quantum numbers, and $m$ are magnetic quantum numbers. $l_3$ satisfies the relation $|l_1 - l_2| \leq l_3 \leq l_1 + l_2$, and the parity of $\bz$ is given by $p(\bz) = p(\bx)p(\by)$.
When building neural networks, the tensor product operation is usually supplemented with an equivariant linear operation to mix channels of each irreducible representation
\begin{align}
    \bz_{c}^{(l_3)} = \sum_{c'} W_{cc'} \bz_{c'}^{(l_3)} \, ,
\end{align}
where $W$ is a trainable weight matrix. We will use the $\otimes$ operator to denote the weighted tensor product operation in the rest of this paper for better readability.

The network structure we use in this work is similar to the Tensor Field Network \cite{Thomas_2018} and NequIP \cite{Batzner_2022}. 
In each message passing layer, the features on each node are updated by collecting information from all its neighbors
\begin{align}
    &\bbf_i^{\prime} = \sum_{j\in\text{neigh}(i)} \bbf_j \otimes  \be_{ij}  \, , \\
    &\be_{ij} = R \left( B( \|\bx_{i j} \| ) \right) Y (\bx_{i j} / \|\bx_{i j} \|) \, . \label{eq:edge}
\end{align}
Here $Y(\bx_{i j} / \|\bx_{i j} \|)$ denotes spherical expansion of the direction of distances between different nodes, $R$ denotes a multi-layer perceptron, and $B$ is a trainable edge length embedding layer as described in Ref.~\onlinecite{Batzner_2022}
\begin{align}
    B(x) = \frac{2}{x_c} \frac{\sin (\frac{b \pi}{x_c} x)}{x} f( x, x_c ) \, ,
\end{align}
where $b$ is a trainable parameter, $x_c$ is a given cut-off length, and $f$ is a polynomial envelop function defined in Ref.~\onlinecite{Gasteiger_2020}.
An equivariant non-linear activation function \cite{Geiger_2022} is applied to all the node features after the updates. 
Note that the spherical expansion of inter-atomic distances fits into the construction of atomic basis functions such as Gaussian type orbitals or Linear Combination of Atomic Orbitals (LCAO), whose angular components are spherical harmonics. 
Since Eq.~\ref{eq:edge} takes relative distances between atom pairs, this type of construction is equivariant with respect to the $E(3)$ group which comprises translations, rotations, and reflections \cite{Geiger_2022}.

The full matrix in atomic orbital space is constructed in a block-wise manner where each block corresponds to the interaction between two atoms.
To obtain these pair-wise features, we use a pair interaction layer to get diagonal and off-diagonal features as in PhiSNet\cite{Unke_2021} and QHNet\cite{Yu_2023}
\begin{subequations}
\begin{align}
    \bbf_{ii} &= \tilde{\bbf}_i + \text{ResBlock}(\tilde{\bbf}_i \otimes \tilde{\bbf}_i) \, ,
    \\
    \bbf_{ij} &= \tilde{\bbf}_i + \text{ResBlock}(\tilde{\bbf}_i \otimes \be_{ij} \otimes \tilde{\bbf}_j) \, .
\end{align}
\end{subequations}
The features $\tilde{\bbf}$ are computed from the node features $\bbf$ via a ResBlock that contains equivariant linear and activation functions
\begin{align}
    \tilde{\bbf} = \text{ResBlock}(\bbf)
    = \text{Linear}(\bbf + \text{Linear}(\text{Activation}(\bbf))) \, .
\end{align}
Each matrix block in the full matrix is constructed using the inverse operation of the tensor product
\begin{align}
    \mathbf{M}_{ij}^{l_1, l_2} = \sum_{l_3=|l_2-l_1|}^{l_2+l_1} \stackrel{l_1, l_2}{\overline{\otimes}} \bbf_{ij}^{(l_3)} .
\end{align}
%

\subsection{Predicting Green's functions and self-energies with equivariant neural networks} \label{sec:gf_nn}

We aim to predict both the static quantity $\bSigma^\text{(HF)}$ (or the Fock matrix $\bF$) and the dynamical quantities $\bG$ and $\tilde{\bSigma}$ using a neural network.
In finite temperature theories, $\bG(i\omega_n)$ and $\tilde{\bSigma}(i\omega_n)$ are functions of imaginary time or Matsubara frequency. Both functions are strongly constrained by their analytical properties, and respecting these properties guarantees, among others, causality and the conservation of probability density \cite{Fei_2021a,Fei_2021b,farid2021luttingerward}. 

To construct such functions, we start from a Lehmann representation
\begin{align}
    &G_{ij}(z) = \frac{1}{Z} \sum_{mn} \frac{\langle n | c_{i} | m \rangle \langle m | c_{j}^{\dagger} | n \rangle}{z + E_n - E_m} (e^{-\beta E_n} + e^{-\beta E_m}) \, , \label{eq:G_lehmann} 
    \\
    &\tilde{\Sigma}_{ij}(z) = \sum_s \frac{t_{is} t_{js}^{*}}{z - t_{ss}} \, , \label{eq:Sigma_lehmann}
\end{align}
where $z$ is a complex frequency value defined on the upper half plane $z \in \mathcal{C}^{+}$, and $Z = \sum_n e^{-\beta E_n}$ is the partition function. In the Lehmann representation of $\tilde{\Sigma}$, $s$ represents virtual orbitals in addition to the physical orbitals, and $t_{xy}$ are terms in a corresponding effective Hamiltonian. See Ref.~\onlinecite{Gramsch_2015} for a detailed derivation. 

The Lehmann representation implies
that  both $\bG(z)$ and $\tilde{\bSigma}(z)$ are Carath\'eodory functions up to a conventional factor of the imaginary unit $i$ \cite{Fei_2021b}.
This mathematical property constrains the values that the functions can assume in the complex plane: Given a set of frequency-dependent values $\bG(z)$ or $\tilde{\bSigma}(z)$, a generalized Pick criterion states that the so-called generalized Pick matrix should be positive semi-definite \cite{Fei_2021b}. Additionally the behavior of $\bG(z)$ and $\tilde{\bSigma}(z)$ for $z\rightarrow i\infty$ is constrained by the short-time evolution of the Hamiltonian and, via the Hamburger moment problem \cite{Akhiezer_1965}, defines the moments of the spectral function.
Values of $\bG$ or $\tilde{\bSigma}$ in the complex plane, and in particular on the real or the imaginary axis, can therefore not be considered as independent quantities and should  not be predicted independently.
The issue can be circumvented by using the Lehmann representations of $\bG$ and $\tilde{\bSigma}$ (Eqs.~\ref{eq:G_lehmann} and \ref{eq:Sigma_lehmann}) directly, which share a general form \cite{Gramsch_2015,Fei_2021b,Huang_2023}
\begin{align}
    \bY(z) = \sum_l \frac{\bX_l}{z - \lambda_l} \, ,
\end{align} 
with $\bX_l$ positive semi-definite (PSD) matrices and $\lambda_l$ real numbers.
Predicting PSD matrices $\bX_l$ associated with real frequency sampling points $\lambda_l$ ensures that the resulting Green's function and self-energy fulfill the Carath\'eodory constraint by construction.

To obtain universal real frequency grids that are applicable to all systems and that scale well as temperature is lowered, we employ the discrete Lehmann representation (DLR) \cite{Kaye_2022a,Kaye_2022b}, which is derived from the truncated spectral Lehmann representation of the imaginary time Green's function
\begin{align}
    \bG(\tau) = \int_{-\Lambda}^{\Lambda} K(\tau, \omega) \boldsymbol{\rho}(\omega) d\omega \, .
\end{align}
Here $\Lambda = \beta \omega_\text{max}$ is a finite truncation parameter, $\boldsymbol{\rho}(\omega)$ is the spectral density, and the analytical continuation kernel is defined as
\begin{align}
    K(\tau, \omega) = - \frac{e^{-\omega \tau}}{1 + e^{-\beta \omega}} \, .
\end{align}
The DLR frequencies $\omega_k$ are chosen based on the discretization of $K(\tau, \omega)$, such that within given accuracy,
\begin{align}
    \bG(\tau) \approx \sum_{l=1}^N K(\tau, \omega_l) \mathbf{g}_l \, ,
\end{align}
and the corresponding spectral function is given by
\begin{align}
    \boldsymbol{\rho}(\omega) = \sum_{l=1}^N \mathbf{g}_l \delta(\omega - \omega_l) \, .
\end{align}
The dynamical self-energy $\tilde{\bSigma}$ follows similar expression as $\bG$.
See Ref.~\onlinecite{Kaye_2022a} for further derivations and additional references.

\subsection{Work flow} \label{sec:workflow}

\begin{figure}[tbh]
    \centering
    \includegraphics[scale=0.22]{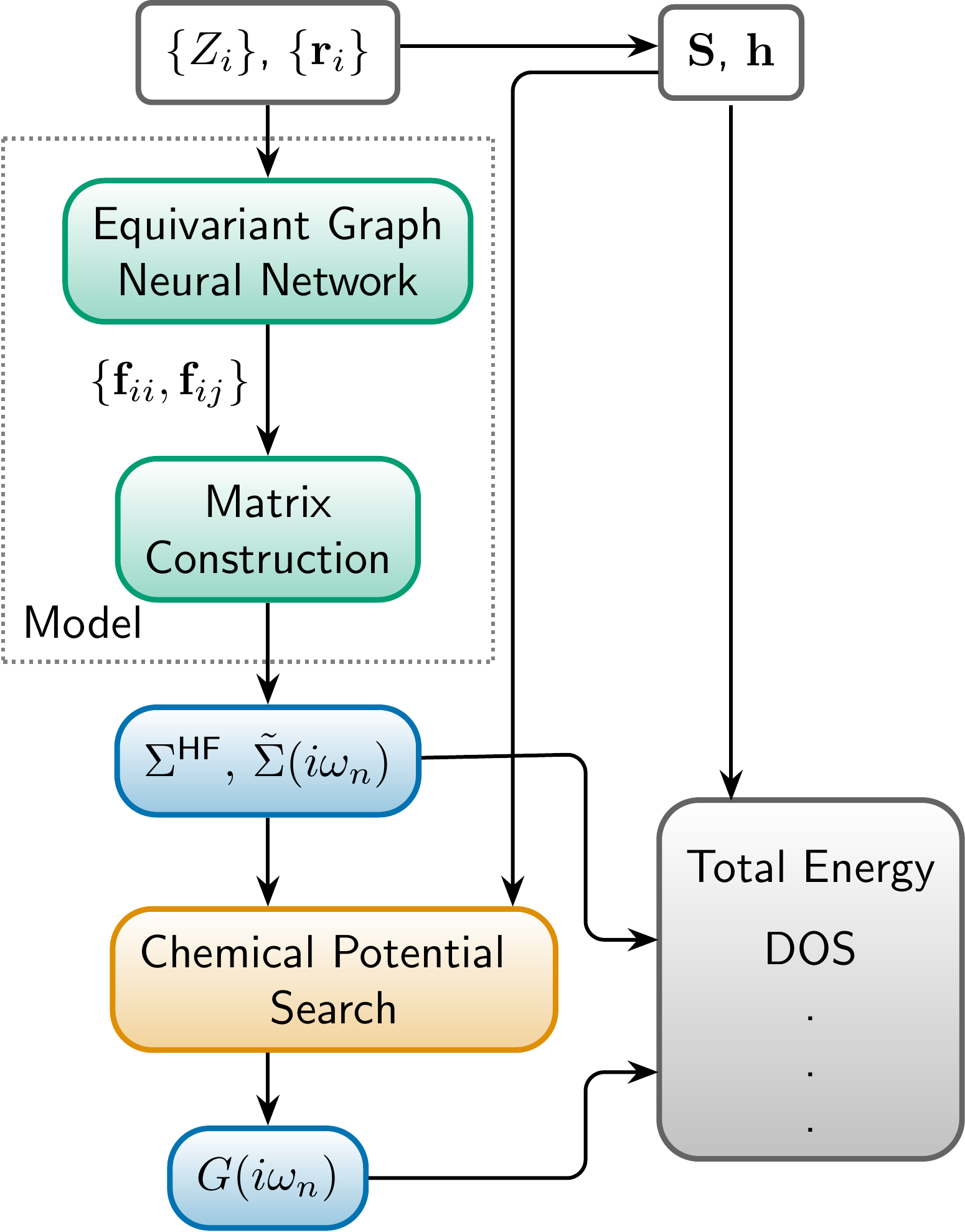}
    \caption{Flow chart of predicting the Green's function and downstream physical observables from atomic charges $\{Z_i\}$ and positions $\{\bbr_i\}$. 
    The green boxes indicate the neural network model with trainable parameters.
    The yellow box specifies the $\mu$-search procedure in Eq.~\ref{eq:dyson_exp}.
    Various physical observables that can be computed from the predicted self-energy and Green's function (blue boxes) are listed in the gray box.
    }
    \label{fig:workflow}
\end{figure}

Fig.~\ref{fig:workflow} demonstrates the general workflow of our method.
In finite temperature grand canonical ensemble calculations, the values of the Green's function are subject to a strict constraint in order to give the correct total number of electrons of the system $N_e = \text{Tr}[\bP \bS]$.
Therefore, we choose to use the neural network as a self-energy solver to predict $\bSigma^\text{(HF)}$ and $\tilde{\bSigma}(i\omega_n)$ instead of predicting $\bG(i\omega_n)$ directly. 
All Matsubara frequency dependent quantities are sampled on the sparse sampling grid \cite{Li_2020} to assure accurate and efficient transformations between the time and frequency domain. 

Starting from the nuclear charges $Z$ and positions $\mathbf{r}$ of atoms, we construct $N+1$ matrices $\bM$ using the equivariant message-passing neural network explained in section \ref{sec:equiv_mpnn}, where $N$ is the number of DLR frequencies used for constructing $\tilde{\bSigma}(i\omega_n)$.
The predicted matrices are symmetrized to ensure the Hermitian and PSD properties needed for constructing $\bSigma^\text{(HF)}$ and $\tilde{\bSigma}(i\omega_n)$
\begin{align}
    &\bM^\text{(Herm)} = \bM + \bM^{\dagger} \, , \\
    &\bM^\text{(PSD)} = \bM \bM^{\dagger} \, .
\end{align}
During the training process of the neural network, the loss is computed by summing over the square of Frobnius norms of matrix differences in $\bSigma^\text{(HF)}$ and $\tilde{\bSigma}(i\omega_n)$ on all frequency points of all data points $x$
\begin{align}
    &l_\text{tot} = \sum_x {l_{x, 1}} + {l_{x, 2}} \, , \\
    &l_{x, 1} = || \bSigma^\text{(HF)}_{x,\text{pred}} - \bSigma^\text{(HF)}_{x,\text{label}} ||^2_F \, , \\
    &l_{x, 2} = \sum_n ||  \tilde{\bSigma}_{x,\text{pred}}(i\omega_n) - \tilde{\bSigma}_{x, \text{label}}(i\omega_n) ||^2_F \, . \label{eq:loss}
\end{align}

With the predicted self-energies, the Green's function can be computed using the Dyson equation (see Eq.~\ref{eq:dyson})
\begin{align}
    \bG(i\omega_n) = \frac{1}{(i\omega_n+\mu)\bS - \bh - \bSigma^\text{(HF)} - \tilde{\bSigma}(i\omega_n)} \, ,
    \label{eq:dyson_exp}
\end{align}
with the chemical potential $\mu$ determined through a chemical potential search procedure such that the total number of electrons matches the target value. The overlap matrix $\bS$ and the one-body integral $\bh$ are easy to compute with given atomic structures and basis functions so we treat them as input.

With the predicted $\bSigma^\text{(HF)}$, $\tilde{\bSigma}(i\omega_n)$, $\bG(i\omega_n)$, we have full access to the one-particle properties of the given electron system. We will show and compare the predicted total energies (Eq.~\ref{eq:e_tot}) and spectral functions (DOS) (Eq.~\ref{eq:DOS}) of different systems in this manuscript.

\section{Results} \label{sec:results}

To demonstrate that our method is general, we benchmark the neural network self-energy solver using both small molecules and periodic systems with different temperatures, Matsubara frequency grids and self-energy approximations.
The results for molecules are obtained at relatively high temperature using the self-consistent second-order Green’s function perturbation theory (GF2) \cite{Dahlen_2005,Phillips_2014,Phillips_2015,Welden_2016,Rusakov_2016} with a Chebyshev sparse sampling grid \cite{Gull_2018,Li_2022}. The convergence to the zero temperature limit is achieved by requiring that the total energy differences between the finite temperature and ground state Hartree Fock (HF) calculations are below $10^{-10}$ Hartree.
For periodic systems the calculations are performed at lower temperature using the self-consistent GW approximation \cite{Yeh_2022} with the intermediate representation (IR) sparse sampling grid \cite{Shinaoka_2017,Chikano_2018,Chikano_2019,Li_2022}, which has a better scaling as a function of temperature.
The GF2 calculations are performed with the full interaction tensor, while the GW calculations are performed with the decomposed interaction using def2-svp-ri auxiliary basis. See appendix \ref{app:diagrammatic} for explicit equations for computing the GF2 and GW self-energy. The interaction tensor, overlap matrix $\bS$ and one-body integral $\bh$ are all generated using the PySCF \cite{Sun_2018,Sun_2020} package.
The DLR frequencies are generated using LibDLR\cite{Kaye_2022b} with $\omega_\text{max} = 100$ for all temperatures. 
Diagrammatic calculations are performed using the Green \cite{Green} open source software package.

In the graph neural network setup, molecules are treated as fully connected graphs, i.e., all atom pairs are connected by edges.
A cut-off radius $r_\text{max}$ is set for periodic systems such that each atom is only connected to all other atoms within this range, see section 
\ref{sec:pbc} for detailed explanations. 
We use a three-layer message passing network for all systems. 

In the post processing procedure, $A(\omega)$ is computed from $\bG(i\omega_n)$ using the Nevanlinna analytical continuation method \cite{Fei_2021a,Green} which is good at resolving sharp peaks around the Fermi level. Each diagonal component of $\bG(i\omega_n)$ in orthogonal basis is continued separately with a broadening parameter $\eta$, and $A(\omega)$ is scaled with $1/(\pi\eta)$ in all plots. 
We use Hartree as energy unit through out the paper with all other quantities presented in units that are in correspondence with the energy unit.

\subsection{Molecules} \label{sec:mol}

\subsubsection{Single water molecule}
As a first proof-of-concept application of the neural network self-energy solver, we compute the Green's functions, total energies, and spectral functions of single water molecule using both the sto-3g and cc-pvdz basis.
All results are obtained at $\beta = 100$ with 320 Chebyshev sparse sampling points.
The configurations are taken from the data set used by Refs.~\onlinecite{Welborn_2018,Cheng_2019}, which are sampled from ab initio molecular dynamics trajectories \cite{Cheng_2019d}.
100 randomly selected configurations are used as test data set, and the rest are used as training data set.

%
\begin{figure}[tbh]
    \centering
    \includegraphics[scale=1.0]{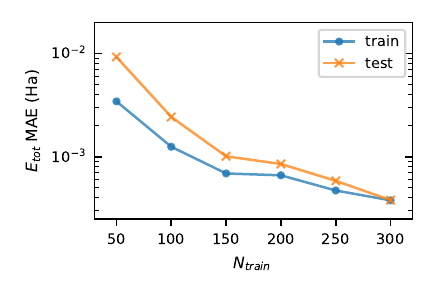}
    \caption{Learning curve for single water molecule with sto-3g basis in terms of the MAE of total energy.
    }
    \label{fig:water_conv}
\end{figure}
%

%
\begin{center}
\begin{table}[tbh]
\begin{tabular}{@{}cccccccc}
\multicolumn{8}{@{}r@{}}{\footnotesize{$\times 10^{-3}$}} \\
\toprule
&$\bSigma^\text{(HF)}$ & $\tilde{\bSigma}(i\omega_n)$ &  $\bG(i\omega_n)$ & $E_\text{tot}$ & HOMO & LUMO & gap \\ [0.5ex] 
\midrule
\multicolumn{7}{@{}l@{}}{sto-3g} \\
&0.0312 & 0.00335 & 0.0971 & 0.380 & 0.163 & 0.251 & 0.361 \\
\midrule
\multicolumn{7}{@{}l@{}}{cc-pvdz} \\
&0.0371 & 0.00320 & 0.190 & 0.819 & 0.162 & 0.265 & 0.253\\
\bottomrule
\end{tabular} \\[0.2cm]
\caption{MAE of 100 testing data for single water molecule.}
\label{tab:water}
\end{table}
\end{center}
%

\begin{figure}[bth]
    \centering
    \includegraphics[scale=1]{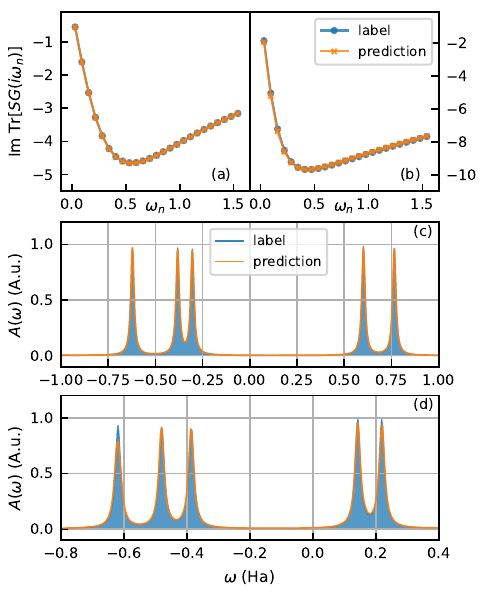}
    \caption{Comparisons of the label and predicted Green's function ((a), (b)) and DOS((c), (d)) of single water molecule. 
    (a, c): Data point with largest MAE in $\bG(i\omega_n)$ with sto-3g basis.
    (b, d): Data point with largest MAE in $\bG(i\omega_n)$ with cc-pvdz basis.
    }
    \label{fig:water_dos}
\end{figure}

Fig.~\ref{fig:water_conv} shows the improvement of accuracy in terms of the mean absolute error (MAE) of total energy in sto-3g basis as we increase the training set size. 
With 100 training data, both the training and testing errors are below 1 mHa per atom, and the train-test gap vanishes with 300 training data.

The testing MAE of various predicted quantities with 300 training data are summarized in Tab.~\ref{tab:water}.
The values of HOMO, LUMO and gap are obtained from the Nevanlinna analytical continuation with $\eta = 0.01$ and a resolution of $10^{-4}$.
Fig.~\ref{fig:water_dos} shows the comparisons of $\text{Tr}[\bS\bG(i\omega_n)]$ and DOS for data points with the largest MAE in $\bG(i\omega_n)$ in the test data sets.
The results presented in Tab.~\ref{tab:water} and Fig.~\ref{fig:water_dos} demonstrate that our method consistently gives accurate estimations for both basis sets. 

\subsubsection{Small organic molecules}

We further benchmark our method with two small organic molecules benzene and ethanol taken from the original MD-17 \cite{Chmiela_2017} data set. 
We trained our model with 800 randomly selected configurations and used an additional 100 randomly selected configurations as test data set for each molecule.
All calculations are performed at $\beta = 100$ with 320 Chebyshev sparse sampling points using the sto-3g basis.

%
\begin{center}
\begin{table}[tbh]
\begin{tabular}{@{}cccccccc}
 \multicolumn{8}{@{}r@{}}{\footnotesize{$\times 10^{-3}$}} \\
 \toprule
 &$\bSigma^\text{(HF)}$ & $\tilde{\bSigma}(i\omega_n)$ &  $\bG(i\omega_n)$ & $E_\text{tot}$ & HOMO & LUMO & gap \\ [0.5ex] 
 \midrule
 \multicolumn{7}{@{}l@{}}{benzene} \\
 &0.158 & 0.00926 & 0.225 & 7.23 & 1.97 & 1.12 & 2.11 \\
 \midrule
 \multicolumn{7}{@{}l@{}}{ethanol} \\
 &0.612 & 0.0622 & 3.19 & 9.54 & 3.35 & 6.01 & 7.96 \\
 \bottomrule
\end{tabular} \\[0.2cm]
\caption{MAE of 100 testing data for benzene and ethanol.}
\label{tab:md-17}
\end{table}
\end{center}
%

\begin{figure}[tbh]
    \centering
    \includegraphics[scale=1]{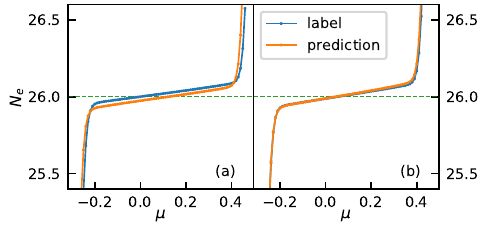}
    \caption{$\mu-N_e$ curve of ethanol with (a) largest MAE in $\bG(i\omega_n)$, (b) smallest MAE in $\bG(i\omega_n)$.}
    \label{fig:ethanol_mu}
\end{figure}

\begin{figure}[tbh]
    \centering
    \includegraphics[scale=1]{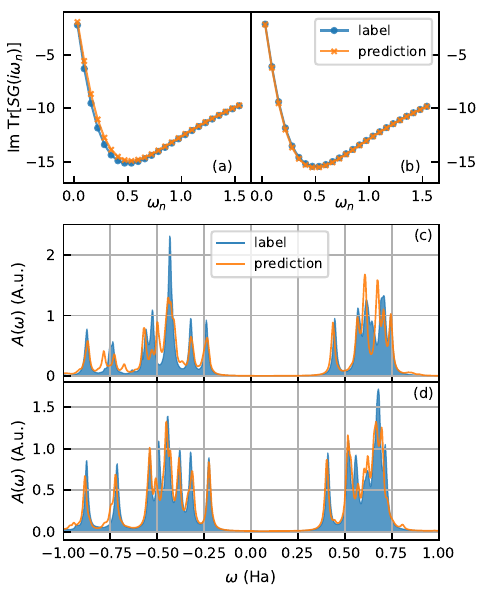}
    \caption{Comparisons of the label and predicted Green's function ((a), (b)) and DOS((c), (d)) of ethanol. 
    (a, c): Data point with largest MAE in $\bG(i\omega_n)$.
    (b, d): Data point with smallest MAE in $\bG(i\omega_n)$.}
    \label{fig:ethanol_dos}
\end{figure}

Table \ref{tab:md-17} summarizes the prediction errors of the two molecules. The values of HOMO, LUMO and gap are obtained from the Nevanlinna analytical continuation with $\eta = 0.01$ and a resolution of $10^{-3}$.
Comparing the results in Tables.~\ref{tab:water} and \ref{tab:md-17}, we see that the prediction errors of benzene and ethanol are larger than those of single water molecule.
This can be primarily attributed to the more complicated atomic configurations of these organic molecules, and the prediction accuracy of benzene is slightly better than ethanol due to its more rigid structure.
Analogous occurrences have been observed in other machine learning models, such as in Refs.~\onlinecite{Chmiela_2017, Welborn_2018, Yu_2023}.
Moreover, as the required matrices become larger in orbital space,
the prediction task becomes more difficult.
As we are only predicting the self-energies, $\bG(i\omega_n)$ and $E_\text{tot}$ are derived properties that do not factor into the supervised learning procedure.
With the current setup, we manage to control the MAE of total energy to around 1 mHa per atom, and the errors of HOMO, LUMO and gap are also at the order of $10^{-3}$.
For applications demanding higher energy accuracy, it is possible to supplement the current workflow with a fine tune procedure utilizing energy as the learning target for better results.

Part of the error amplification from $\bSigma$ to $\bG$ and $E_\text{tot}$ comes from the chemical potential search procedure introduced in section \ref{sec:workflow}. Fig.~\ref{fig:ethanol_mu} shows the $\mu-N_e$ curve of data points with the largest and smallest MAE in $\bG(i\omega_n)$ for ethanol. 
Since the $\mu-N_e$ curve is relatively flat around the target electron number, small errors in $\bSigma^\text{HF}$ and $\tilde{\bSigma}(i\omega_n)$ might cause a non-negligible $\mu$-shift that propagates to $\bG(i\omega_n)$. 
The comparison of $\text{Tr}[\bS\bG(i\omega_n)]$ and DOS of these two data points are shown in Fig.~\ref{fig:ethanol_dos}. 
For data point with the largest $\bG$ error, the shift in $\mu$ causes an obvious shift in the low frequency part of $\bG(i\omega_n)$ as shown in panel(a). However, this shift would not significantly affect the band gap or HOMO LUMO as shown in panel (c) since the frequency grid is shifted by $\mu$ in the mean time.

\subsection{Periodic system} \label{sec:pbc}

\begin{figure}[tbh]
    \centering
    \includegraphics[scale=0.5]{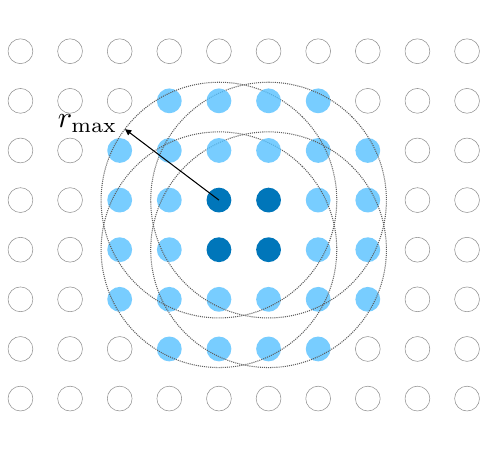}
    \caption{Schematic plot for periodic system set up. 
    Dark blue dots represent atoms in the center unit cell, dotted lines indicate the cut-off range, and light blue dots are the considered images of atoms.
    }
    \label{fig:pbc}
\end{figure}

The neural network self-energy solver can be applied to periodic systems in a similar way as isolated molecules.
For periodic system calculations carried out in k-space, the matrix elements are determined by summing the contributions from all unit cells, each labeled by $\mathbf{R}$, within the periodic lattice
\begin{align}
    \bM(\bk) = \sum_{\bR} e^{-i \bk \cdot \bR} \, \bM(\bR) \, .
    \label{eq:pbc_sum}
\end{align}
In the corresponding graph neural network setup, each atom in the center unit cell is connected to all atom images within a given cut-off radius $r_\text{max}$. 
This setup employs a local approximation to address the periodicity of the system \cite{Xie_2018,Batzner_2022,Li_2022,Yang_2023}. 
See Fig.~\ref{fig:pbc} for a schematic plot.
Unlike constructing the real space matrices in LCAO which naturally fits into this type of tight-binding setup \cite{Li_2022,Gong_2023,Yang_2023}, 
recovering k space matrices in Gaussian basis requires summing over the features from different unit cells with k-dependent phase factors given in Eq.~\ref{eq:pbc_sum}.

As an example, we apply our method to Gamma point calculations of diamond and silicon using their conventional cell with lattice parameters $3.57 \text{\AA}$ and $5.43 \text{\AA}$. The configurations taken from the data set of Ref.~\onlinecite{Gu_2023}.
100 (200) randomly drawn configurations are used as training data for diamond (silicon) and the models are tested with another 100 randomly selected configurations. $r_\text{max}$ is set to be $6 \text{\AA}$ for diamond and $10 \text{\AA}$ for silicon.
To improve prediction accuracy, we trained two neural networks separately for $\bSigma^\text{(HF)}$ and $\tilde{\bSigma}(i\omega_n)$ in this example. 
All calculations are performed at $\beta = 500$ with 136 IR sparse sampling points generated with $\Lambda = 10^5$ using the sto-3g basis.

%
\begin{center}
\begin{table}[tbh]
\begin{tabular}{@{}cccccccc}
 \multicolumn{8}{@{}r@{}}{\footnotesize{$\times 10^{-3}$}} \\
 \toprule
 &$\bSigma^\text{(HF)}$ & $\tilde{\bSigma}(i\omega_n)$ & $\bG(i\omega_n)$ & $E_\text{tot}$ & HOMO & LUMO & gap\\ [0.5ex] 
 \midrule
 \multicolumn{7}{@{}l@{}}{diamond} \\
 &0.0808 & 0.0118 & 0.507 & 3.36 & 1.09 & 1.21 & 1.55\\
 \midrule
 \multicolumn{7}{@{}l@{}}{silicon} \\
 &0.0933 & 0.00371 & 0.528 & 8.01 & 0.463 & 1.67 & 1.72\\
 \bottomrule
\end{tabular} \\[0.2cm]
\caption{MAE of 100 testing data for diamond and silicon.}
\label{tab:pbc}
\end{table}
\end{center}
%

The prediction MAE are summarized in table \ref{tab:pbc} with the Nevanlinna continuation performed with $\eta = 0.005$ and a resolution of $10^{-3}$.
As illustrated in the table, 
our method demonstrates the ability to predict the total energies and spectral properties of periodic systems with an accuracy comparable to that achieved for molecules,
indicating that this method is promising for more complicated applications in real material calculations.

\section{Conclusion and outlook} \label{sec:conclusion}

In this manuscript, we introduce a general framework for predicting the finite temperature self-energy and Green's function using equivariant neural network. 
The proof of concept examples demonstrate that from the predicted self-energy and Green's function, we are able to obtain fairly accurate energy and band gap values for both molecules and periodic systems.

The inference of the neural network scales quadratically with the number of atoms, which is much more efficient than performing actual many-body calculations. Therefore, a trained model could be used for rapid preliminary calculations of electron systems to identify desired properties. On the other hand, the predicted self-energy and Green's function can also serve as a reasonable initial guess of corresponding self-consistent diagrammatic method which accelerates the convergence.
For future developments, including specific observables such as energy in the loss function Eq.~\ref{eq:loss} is expected to give improved accuracy when accurate results are required.
Integrating recent developments of equivariant graph neural network \cite{Musaelian_2023,Yu_2023,Passaro_2023} into our model is anticipated to further improve the prediction accuracy and efficiency.
Besides observables that are directly related to the Green's function and self-energy, force and other response properties would also be accessible via automatic differentiation through the trained model.

In summary, the equivariant neural network self-energy solver provides a new opportunity to leverage the rapid development of geometric deep learning to fast and accurate prediction of molecular and material properties at many-body level. 

\begin{acknowledgements}

We would like to acknowledge Yixiao Chen and Zuxin Jin for useful discussions, Sergei Iskakov for helping with electronic structure calculations, and Hugo U.R. Strand for providing feedback on the manuscript. L.W. was supported by the National Natural Science Foundation of China under Grants No. T2225018, No. 92270107, No. 12188101, and No. T2121001, and the Strategic Priority Research Program of Chinese Academy of Sciences under Grants No. XDB0500000 and No. XDB30000000. E.G. was supported by NSF OAC Grant no. 2310582.
\end{acknowledgements}

\appendix

\section{Self-consistent Diagrammatic methods} \label{app:diagrammatic}

This appendix provides the equations for approximating the self-energy within self-consistent diagrammatic methods.
The orbital index $i$ and spin index $\sigma$ will be written separately for clarity.

The HF self-energy is static (frequency independent) and only depends on the density matrix $\bP$
\begin{align}
    &\Sigma^\text{(HF)}_{ij, \sigma} = \sum_{kl}\sum_{\sigma'} (U_{ijkl} - U_{ilkj}\delta_{\sigma \sigma'}) P_{kl, \sigma'} \, , \\
    &P_{kl, \sigma} = G_{lk, \sigma}(\tau = 0^{-}) \, .
\end{align}
In the GF2 approximation, the dynamical part of $\bSigma$ is approximated with two second-order bold self-energy diagrams. The corresponding second order self-energy is given by
\begin{align}
    \Sigma^\text{(GF2)}_{ij, \sigma}&(\tau) = -\sum_{klmnpq} U_{ilnp}G_{lk, \sigma}(\tau) \label{eq:sigma_GF2} \\
    &\times \sum_{\sigma'} G_{pq, \sigma'}(\tau) G_{mn, \sigma'}(-\tau) (U_{kjqm} - U_{qjkm}\delta_{\sigma \sigma'}) \, . \nonumber
\end{align}
Within the self-consistent GW approximation, the dynamical part of the self-energy consists of an infinite series of RPA-like bubble diagrams. The self-energy $\bSigma^\text{(GW)}$ reads as
\begin{align}
  \Sigma_{ij,\sigma}^{(\text{GW})}&(\tau) = -\sum_{kl} \tilde{W}_{ilkj}(\tau) G_{lk,\sigma}(\tau) \, ,
\end{align}
where $\tilde{\mathbf{W}}$ is the effective screened interaction.
The GW self-energy is usually computed using decomposed interaction to get a better scaling, i.e., the interaction tensor is written in a decomposed form
\begin{align}
    U_{ijkl} = \sum_Q V_{ij}(Q) V_{kl}(Q) \, ,
\end{align}
and the effective screened interaction is given by
\begin{align}
    \tilde{W}_{ijkl}(i\Omega_n) = \sum_{QQ'} V_{ij}(Q) P_{QQ'}(i\Omega_n) V_{kl}(Q') \, ,
\end{align}
with $\Omega_n = 2n\pi/\beta$, $n \in \mathbb{Z}$ the bosonic Matsubara frequencies. $\bP$ is an auxiliary function given by
\begin{align}
    &\bP(i\Omega_n) = [\mathbf{I} - \bP_0(i\Omega_n)]^{-1} \bP_0(i\Omega_n) \, , \\
    &\bP_{0, QQ'}(\tau) =  \\
    & -\sum_{\sigma \sigma'} \sum_{abcd} V_{da}(Q) \times G_{c\sigma',d\sigma}(-\tau) G_{a\sigma, b\sigma'}(\tau) V_{bc}(Q') \, . \nonumber
\end{align}

\bibliography{refs}

\clearpage
\end{document}